\documentclass[12pt]{article}
\usepackage{amssymb}
\usepackage{graphicx}
\usepackage[cp866]{inputenc}

\begin{document}
 \begin{center}

{\bf Perturbative and non-perturbative effects \\ in ultraperipheral production
of lepton pairs }

\vspace{2mm}

I.M. Dremin$\footnote{dremin@lpi.ru}$

{\it Lebedev Physical Institute, Moscow, Russia}

\end{center}


\vspace{1mm}

\begin{abstract}
Perturbative and non-perturbative terms of the cross sections of
ultraperipheral production of lepton pairs in ion collisions are taken into 
account. It is shown that production of low-mass $e^+e^-$ pairs 
is strongly enhanced (compared to perturbative estimates)
due to the non-perturbative Sommerfeld-Gamow-Sakharov (SGS) factor. 
Coulomb attraction of the non-relativistic components of those pairs leads
to the finite value of their mass distribution at lowest masses. 
Their annihilation can result in the increased intensity of 511 keV photons. It 
can be recorded at the NICA collider and is especially crucial in astrophysical 
implications regarding the 511 keV line emitted from the Galactic center.
The analogous effect can be observed in lepton pairs production at LHC.
Energy spectra of lepton pairs created in ultraperipheral nuclear collisions
and their transverse momenta are calculated.
\end{abstract}

\vspace{1mm}


\vspace{1mm}
Interaction of two photons from electromagnetic fields of colliding ions
can lead to production of lepton pairs. They are produced in grazing 
ultraperipheral collisions. This process was first treated perturbatively
by Landau and Lifshitz in 1934 \cite{lali}. It was shown that its total 
cross section rapidly increases with increasing energy $E$ as $\ln ^3E$ 
in asymptotics. This is still the strongest asymptotical
energy dependence known in particle physics. Moreover, the numerical factor 
$Z^4\alpha ^4$ in the total cross section compensates the effect of the small 
electromagnetic coupling $\alpha $ for heavy ions with large charge $Ze$. 
Therefore, the ultraperipheral production of $e^+e^-$-pairs (as well as 
$\mu ^+\mu ^-$ etc) in ion collisions can become
the dominant mechanism at very high energies. It is already widely studied
at colliders. The heuristic knowledge of these processes is helpful in
understanding some important astrophysical phenomena as well.

Even earlier, almost a century ago, in 1924, Fermi \cite{fer1, fer2} considered 
the general problem of interaction of charged objects with matter: 
{\it "Let's calculate, first of all, 
the spectral distributions corresponding to those of the electric field 
created by a particle with electric charge, $e$, passing with velocity, $v$, 
at a minimum distance, $b$, from a point, $P$."} Fermi obtained the formula for 
the intensity of the electromagnetic field created by a moving charged object. 
It was used in 1934 by Weizs\"acker \cite{wei} and Williams \cite{wil} for their 
formulation of the method of equivalent photons widely applied nowadays for 
treatment of the ultraperipheral interactions. Solutions of Maxwell equations
for the electromagnetic fields of the moving charge are used to derive
the Poynting vector. The component of the Poynting vector along the direction 
of colliding electromagnetic charges defines the total energy of equivalent 
photons and their energy distribution.

The~distribution of equivalent photons with a fraction of the nucleon energy
$x$ generated by a moving nucleus with the charge $Ze$ can be obtained 
from the expression for the Poynting vector as
\begin{equation}
\frac {dn}{dx}=\frac {2Z^2\alpha }{\pi x}\ln \frac {u(Z)}{x}
\label{flux}
\end{equation}
if integrated over the transverse momentum of scattered nuclei up to some value 
(see, e.g., Ref. \cite{blp}). The physical meaning of the ultraperipherality
parameter $u(Z)$ is the ratio of the maximum adoptable transverse momentum
to the nucleon mass as the only massless parameter of the problem.
Its value is determined by the form factors of colliding ions responsible
for their entity (see, e.g., \cite{vyzh}).
It is clearly seen from Eq. (\ref{flux}) that soft photons with small
fractions $x$ of the nucleon energy dominate in these fluxes.

Pairs of leptons with opposite charges are produced in grazing collisions of 
interacting ions where two photons from their electromagnetic clouds interact. 
Abundant creation of pairs with rather low pair-masses is the typical feature 
of ultraperipheral interactions \cite{20dr}. Two-photon fusion production of 
lepton pairs has been calculated with both the equivalent photon approximation 
proposed in \cite{wei, wil} and via full lowest-order QED calculations
\cite{lali, rac, bgms, sergey1, sergey2} reviewed recently in \cite{drufn}.
According to the equivalent photon approximation, the general expression for
the total cross section looks like
\begin{equation}
\sigma _{up}(X)=
\int dx_1dx_2\frac {dn}{dx_1}\frac {dn}{dx_2}\sigma _{\gamma \gamma }(X).
\label{e2}
\end{equation}
Feynman diagrams of ultraperipheral processes contain the subgraphs of 
two-photon interactions leading to production of some final states $X$ 
(e.g., $e^+e^-$ pairs). These 
blobs can be represented by the cross sections of the corresponding processes.
Therefore, $\sigma _{\gamma \gamma }(X)$ in (\ref{e2}) denotes the total cross
section of production of the state $X$ by two photons from the electromagnetic
clouds surrounding colliding ions and $dn/dx_i$ describe the densities of
photons carrying the share $x_i$ of the ion energy. 
The spectra of lepton pairs created in ultraperipheral collisions can be 
obtained from Eq. (\ref{e2}) by omitting the relevant integrations.

Both bound (e.g., para- and ortho-positronia for $e^+e^-$) and unbound pairs 
can be created. However, the bound pairs are less intensively produced than 
the unbound ones (see, \cite{20dr}).
Therefore we consider here the processes with unbound pairs only.
The cross section usually inserted in (\ref{e2}) in case of creation of the 
unbound pairs $X=e^+e^-$ was calculated by Breit and Wheeler in the 
lowest order perturbative approach and looks \cite{blp,brwh} as
\begin{equation}
\sigma _{\gamma \gamma }^{BW}(X)=\frac {\pi \alpha ^2}{m^2}(1-v^2)
[(3-v^4)\ln \frac {1+v}{1-v}-2v(2-v^2)],
\label{mM}
\end{equation}
where $v=\sqrt {1-\frac {4m^2}{M^2}}$ is the relative velocity of a pair 
component to the pair center of mass, 
$m$ and $M$ are the electron and pair masses,
correspondingly. The Breit-Wheeler cross section tends to 0 at the threshold 
of pair production $M=2m$ ($v=0$) and decreases as $\frac {1}{M^2}\ln M$ at 
very large $M$ ($v\rightarrow 1$).

However, one must take into account the specific attractive long-range Coulomb 
forces acting non-perturbatively between the leptons with opposite charges. 
At the production point, the components of pairs with low masses $M$ close 
to 2$m$ move very slowly relative to one another. They are strongly influenced 
by the attractive Coulomb forces. In the non-relativistic limit, these states 
are transformed by mutual interactions of the components to form effectively 
a composite state whose wave function is a solution of the relevant
Schroedinger equation. The normalization of Coulomb wave functions plays an
especially important role at low velocities. It differs from the normalization
of free motion wave functions used in the perturbative derivation of
Eq. (\ref{mM}). 

The normalization of the unbound pair wave function reads \cite{llqm}
\begin{equation}
|\psi (\vec r=0)|^2 =\frac{\pi\xi}{sh(\pi\xi)}e^{\pi\xi}
=\frac{2\pi\xi}{1-e^{-2\pi\xi}};~~~\xi=\frac {\alpha }{v}.
\label{psi}
\end{equation}
This is the widely used Sommerfeld-Gamow-Sakharov (SGS) factor 
\cite{som, gam, somm, sakh} which is responsible for the non-perturbative 
contribution to the matrix element. It results in the so-called 
"$\frac {1}{v}$-law" of the enlarged outcome of the reactions with extremely 
low pair masses. This factor is described in the standard
textbooks (see, e.g., the 4-th and later editions of the Landau-Lifshitz book
on non-relativistic quantum mechanics \cite{llqm})
and used in various publications (e.g., \cite{baier, ieng, cass, arko}). 
The Sakharov recipe of its account for production of $e^+e^-$-pairs desctibed 
in \cite{sakh} consists in direct multiplication of the Breit-Wheeler 
cross section by the SGS-factor 
\begin{equation}
T=\frac {2\pi \alpha}{v(1-\exp (-2\pi \alpha/v))},
\label{sgs}
\end{equation}
such that the cross section $\sigma _{\gamma \gamma }(X)$ in (\ref{e2}) is
\begin{equation}
\sigma _{\gamma \gamma }(X)=\sigma ^{BW}_{\gamma \gamma }(X)T.
\label{crs}
\end{equation}

The differential distributions of leptons are easily computed from
the integrands of (\ref{e2}). For example, the distribution of the
velocity $v$ of pair' components with account of the SGS-factor 
was first obtained in \cite{dgm}:
\begin{equation}
\frac {d\sigma }{dv^2}=\frac {16Z^4\alpha ^5}{3m^2v}
\frac {(3-v^4)\ln \frac {1+v}{1-v}-2v(2-v^2)}
{1-\exp (\frac {-2\pi \alpha}{v})}
\ln^3 \frac {u\sqrt {s_{nn}(1-v^2)}}{2m}.
\label{sv2}
\end{equation}
Here $\sqrt {s_{nn}}$ is the total energy of two colliding nucleons in the
center of mass system. It can be
represented by the following expressions $s_{nn}=4m^2_n\gamma ^2_c=
2m^2_n(\gamma _r+1)=2m_n(E_k+2m_n)$ where $m_n$ is a nucleon mass,
$\gamma _c$ and $\gamma _r$ are the Lorentz-factors of the nucleon in the 
center of mass and rest (of another nucleon) systems and $E_k$ is
the excess of the total energy of an impinging proton in the rest system
of a target proton over its mass that corresponds to the nucleon kinetic 
energy in the non-relativistic domain. The threshold of $e^+e^-$-pair creation 
in the rest system of one of the nucleons is 
$E_{k,t}=4m(1+\frac {m}{2m_n})\approx 2.05$ MeV.

The distribution (\ref{sv2}) is shown in the left-hand side of Fig. 1. 

\begin{figure}

\centerline{\includegraphics[width=\textwidth]{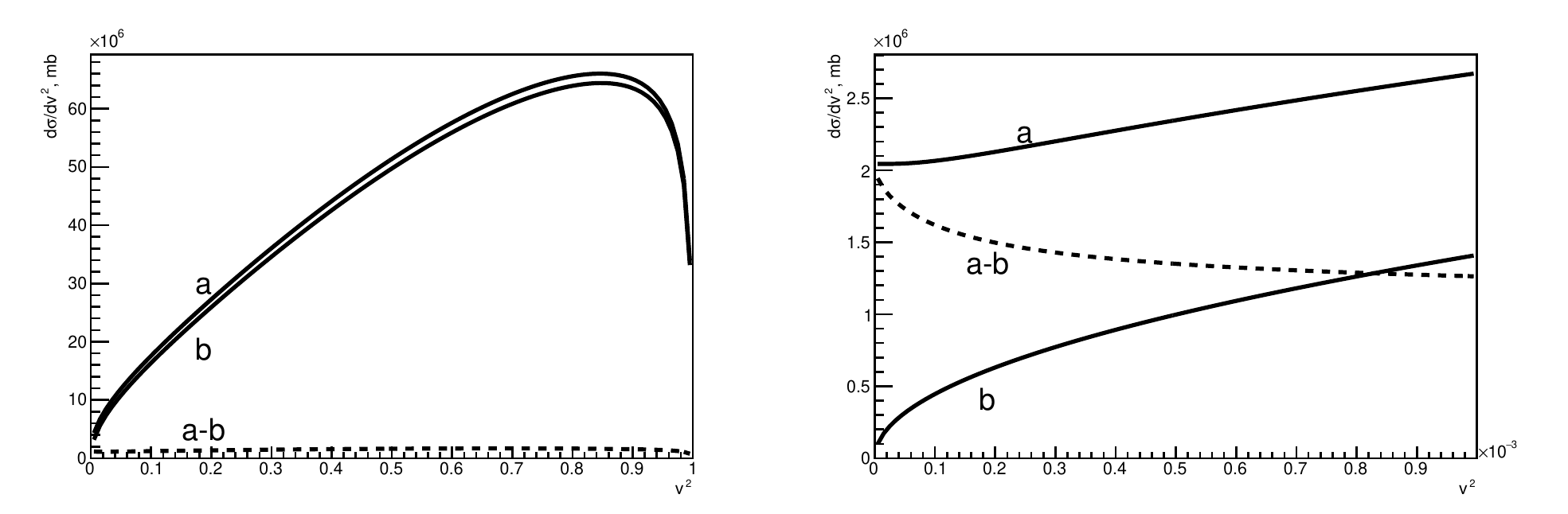}}

Fig. 1. The distribution of the velocities $v$ in $e^+e^-$-pairs produced 
in ultraperipheral Au-Au-collisions at NICA energy $\sqrt {s_{nn}}$=11 GeV
with (a) and without (b) account of the SGS-factor \cite{dgm}. 
For the region of small 
velocities (masses) they are shown in the right-hand side. Their difference 
(a-b) is shown by the dashed line. They tend to a constant at $v \rightarrow 0 $ 
with accounted SGS-factor and to 0 without its account. 

Note the factors $10^{-3}$ at the abscissa scale and $10^6$ at the ordinate.
\end{figure}

Its most interesting feature is not clearly seen in the left-hand side and, 
therefore, it is deciphered at the larger scale in the right-hand side of the 
same Figure for small velocities $v$. It demonstrates the crucial 
difference between the distributions with (a) and without (b) account of the 
SGS factor. At low velocities $v$ (i.e., small masses $M=2m/\sqrt {1-v^2}$) the 
two curves tend to different values. It is finite with account of SGS factor 
and vanishes without it. This is a clear signature of its $1/v$-law.
In general, the non-relativistic nature of the pair of annihilating
particles separates the short-distance annihilation process (taking place at
 distances up to O(1/$m$)) from the long-distance interactions (characterized by
the Bohr radius of the pair O(1/$m\alpha $)), responsible for the SGS-effect.
The same peculiar feature of finite $d\sigma /dM^2$ at $M$ close to $2m$
is seen in the distribution of masses $M$ (see Ref.\cite{dgm}).

The integral contribution of the non-perturbative factor is small 
as discussed in Ref. \cite{dgm} but it is crucial for production of pairs 
with low velocities of their components (see Fig. 1). 
This remarkable effect of the mutual attraction of the created components 
is well known.

Protons are the main component of the ion flows in star explosions.
The energy behaviour of the total cross section of the ultraperipheral 
production of $e^+e^-$-pairs in proton-proton collisions \footnote{The ion-ion 
cross sections are $Z^2_1Z^2_2$-times larger.} is shown in Fig. 2.
It is obtained by the integration of Eq. (\ref{sv2}) over all admissible 
velocities $v$ at a given energy $s_{nn}$. They are defined by the requirement 
to the argument of the logarithm being larger than 1. The cross section does 
not vanish at the very threshold of pair production 
$\sqrt {s_t}=2(m_p+m)\approx 1.88$ GeV but stays finite if the energy 
dependence in (\ref{sv2}) is boldly extended to the threshold. Surely, this 
is an artefact of its extrapolation down to low energies. It must tend to zero 
at the threshold because the small masses close to $2m$ (i.e., small 
velocities $v$) are important near it. At low energies one has to integrate up 
to small velocities. The finiteness of values of $d\sigma /dv^2$ at the very 
threshold is provided by the $1/v$-law of the SGS-factor. Similar to shown 
in Fig. 1, it implies that the total cross section must increase 
linearly with $s-s_t$ there. The non-perturbative factor is most important 
near the threshold.

\begin{figure}

\centerline{\includegraphics[width=\textwidth]{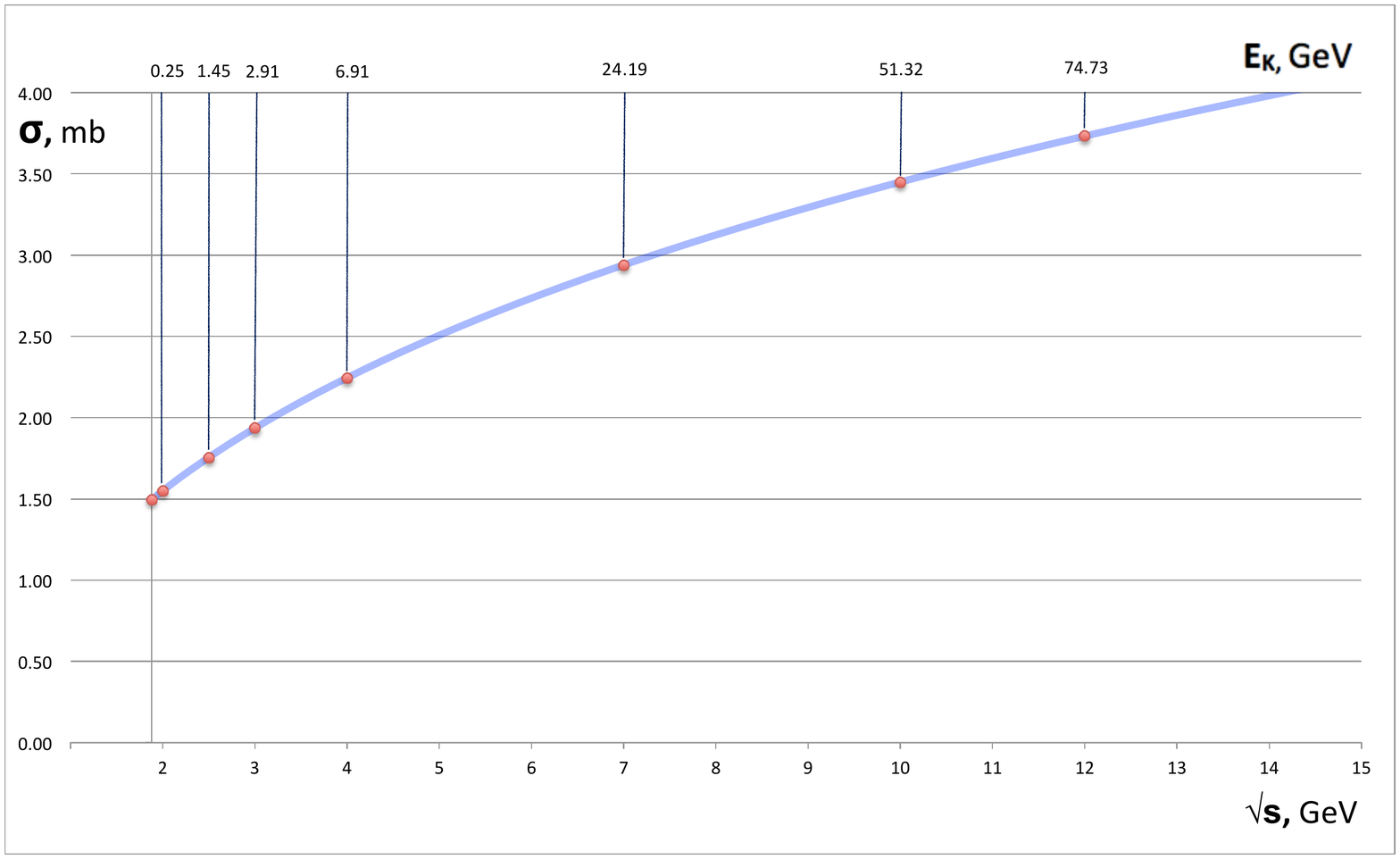}}

Fig. 2. The energy behaviour of the total cross section of ultraperipheral
proton-proton interactions from the threshold of $e^+e^-$-production 
to NICA energies. ($s\equiv s_{nn}$ for $pp$-collisions.)
\end{figure}

The knowledge of values of the ultraperipheral cross sections at low energies 
is especially important for astrophysical applications where the energy 
spectrum of charged hadrons created in star explosions peaks at rather 
low energies. Correspondingly, the upper abscissa axis shows the values
of the relativistically invariant variable $E_k=(s-4(m_p+m)^2)/2m_p$
which coincides with the kinetic energy of an impinging proton in the 
non-relativistic region of collisions treated in the rest system of another 
proton. The total cms-energy of the colliding protons $\sqrt s$ is shown at the
lower abscissa axis.

The flows of created slow electrons and
positrons can become rather high to provoke numerous annihilations and
emission of 511 keV photons. Further knowledge of these flows can be got
from theoretical distributions of energies and transverse momenta
of produced electrons and positrons.
The energy distributions of electrons and positrons created in ultraperipheral
processes coincide with the corresponding distributions of photons in the
clouds around the colliding nuclei. Therefore they are directly obtained from
 Eq. (\ref{e2}) by omitting the $x_i$-integrations. One gets
\begin{equation}
\frac {d^2\sigma }{dE_1dE_2}=\frac {4(Z\alpha )^4}{m^4}
\frac {\alpha (1-v^2)^2[(3-v^4)\ln \frac {1+v}{1-v}-2v(2-v^2)]}
 {v(1-\exp (-\frac {2\pi \alpha}{v}))}
 \ln \frac {u\sqrt {s_{nn}}}{E_1} \ln \frac {u\sqrt {s_{nn}}}{E_2}.
\label{sE}
\end{equation}
As usual, here $v=\sqrt {1-\frac {4m^2}{M^2}}=\sqrt {1-\frac {m^2}{E_1E_2}}$.
The low-energy electrons and positrons are favoured.
 
The transverse momentum $p_t$-distribution of leptons can also be obtained from 
the general formula (\ref{e2}). The non-perturbative effects are prevailing
in the region of small masses. Their integral contribution is rather low.
Therefore, the main features can be seen if the differential Breit-Wheeler 
distribution is inserted there. One gets:
\begin{equation}
\frac {d\sigma ^{BW}}{dp_t}=\frac {2\pi \alpha ^2 (1-v^2)}{m^2p_T}
\frac {1-p^2_t(1-v^2)/2m^2}{\sqrt {1-p^2_t(1-v^2)/m^2}}.
\label{p_t}
\end{equation}
The transverse momenta of leptons are strongly limited in ultraperipheral 
production for $p_t\geq m$ as
\begin{equation}
\frac {d\sigma ^{up}}{dp_t}=\frac {16(Z\alpha )^4}{9\pi p^3_t}
\ln ^3\frac {u^2s_{nn}}{p^2_t}.
\label{sp_t}
\end{equation}

Summarizing, it is shown that the total cross sections and differential
distributions of lepton production in ultraperipheral nuclear collisions
can be calculated with account of both perturbative (Breit-Wheeler) and
non-perturbative (Sommerfeld-Gamov-Sakharov) contributions. The rapid energy
increase and the finite values of the differential cross sections at the 
threshold are the most peculiar features of these processes. Their particular 
values are much larger for heavy nuclei with large charge $Ze$ surrounded by 
stronger electromagnetic fields. It results in the factor $Z^4$ for the  
cross sections of identical ions. Emission of low-mass pairs of leptons is
favoured.

\vspace{6pt}
{\bf Acknowledgments}

I am grateful to V.G. Terziev for help with Fig. 2.

This work is supported by the RFBR project 18-02-40131.


\vspace{6pt}

\begin{thebibliography}{999}
\bibitem{lali}
L.D. Landau, E.M. Lifshitz, Phys. Zeit. Sowjet. {\bf 6},~244 (1934)
\bibitem{fer1}
E. Fermi, Nuovo Cimento {\bf 2}, 143 (1924)
\bibitem{fer2}
E. Fermi, Z. Physik {\bf 29}, 315 (1924) 
\bibitem{wei}
C.F.V. Weizs\"{a}cker, Z. Phys. ~{\bf 88},~612 (1934)
\bibitem{wil}
E.J. Williams, Phys. Rev. ~{\bf 45}, 725 (1934)
\bibitem{blp}
V.B. Berestetsky, E.M. Lifshitz, L.P. Pitaevsky, 
Kvantovaya Electrodinamika
(Fizmatlit: Moscow, Russia, 2001).
\bibitem{vyzh}
M.I. Vysotsky, E.V. Zhemchugov, Phys. Usp. {\bf 189}, 975 (2019)
\bibitem{20dr}
I.M. Dremin, Universe {\bf 6(7)}, 94 (2020)
\bibitem{rac}
G. Racah, Nuovo Cim. {\bf 14}, 93 (1937)
\bibitem{bgms}
V.M. Budnev, I.F. Ginzburg, G.V. Meledin, V.G. Serbo, Phys. Rep. {\bf 15}, 181 
(1975)
\bibitem{sergey1}
E. Bartosh, S.R. Gevorkyan, E.A. Kuraev, N.N. Nikolaev
Phys. Rev. A {\bf 66}, 042720 (2002)
\bibitem{sergey2}
S.R. Gevorkyan, E.A. Kuraev, J. Phys. G {\bf 29}, 1 (2013) 
\bibitem{drufn}
I.M. Dremin,  Phys. Usp. {\bf 190}, 811 (2020) 
\bibitem{brwh}
G. Breit, J.A. Wheeler, Phys. Rev. {\bf 46}, 1087 (1934)
\bibitem{llqm}
L.D. Landau, E.M. Lifshitz, Kvantovaya Machanika, Nerelyativistskaya Teoriya,
(Fizmatlit, Moscow, 1968) (translated in Quantum Mechanics, Non-relativistic 
theory (Pergamon Press, Oxford, 1977))
\bibitem{som}
A. Sommerfeld, Atombau und Spectrallinien (F. Vieweg und Sohn, Brunswick, 
Deutschland, 1921)
\bibitem{gam}
G. Gamow, Z. Phys. {\bf 51}, 294 (1928)
\bibitem{somm}
A Sommerfeld, Ann. Phys. (Leipzig) {\bf 403}, 257 (1931)
\bibitem{sakh}
A.D. Sakharov, Zh, Exp. Teor. Fiz. {\bf 18}, 631 (1948) (Reprinted in
Phys. Usp. {\bf 34}, 375 (1991))
\bibitem{baier}
V.N. Baier, V.S. Fadin, Sov.Phys. JETP {\bf 30}, 127 (1970)
\bibitem{ieng}
R. Iengo, JHEP {\bf 05}, 024 (2009)
\bibitem{cass}
S. Cassel, J. Phys. G {\bf 37}, 105009 (2010)
\bibitem{arko}
A.B. Arbuzov, T.V. Kopylova, JHEP {\bf 04}, 009 (2012)
\bibitem{dgm}
I.M. Dremin, S.R. Gevorkyan, D.T. Madigozhin, EPJC {\bf 81}, 276 (2021)
\end{thebibliography}
\end{document}